\documentclass[conference]{./IEEEtran}

\let\labelindent\relax

\usepackage[normalem]{ulem}
\usepackage{amsmath}
\usepackage{algorithm}
\usepackage{color}
\usepackage{comment}
\usepackage{hyperref}
\usepackage{enumerate}
\usepackage{paralist}
\usepackage{times}
\usepackage{epsfig}
\usepackage{amssymb}
\usepackage{verbatim}
\usepackage{url}
\usepackage{xspace}
\usepackage{grffile}
\usepackage{wrapfig}
\usepackage{sidecap}
\usepackage{microtype}
\usepackage{enumitem}
\usepackage[noend]{algpseudocode}
\usepackage{setspace}
\usepackage{graphicx}
\usepackage{multicol}
\usepackage{subfig}
\usepackage[labelfont=bf, justification=justified]{caption}
\usepackage{pifont}

\newcommand{\pgftextcircled}[1]{
    \setbox0=\hbox{#1}%
    \dimen0\wd0%
    \divide\dimen0 by 2%
    \begin{tikzpicture}[baseline=(a.base)]%
        \useasboundingbox (-\the\dimen0,0pt) rectangle (\the\dimen0,1pt);
        \node[circle,draw,outer sep=0pt,inner sep=0.1ex, fill=black] (a) {#1};
    \end{tikzpicture}
}

\algdef{SE}[DOWHILE]{Do}{doWhile}{\algorithmicdo}[1]{\algorithmicwhile\ #1}
\usepackage[noend]{algpseudocode}
\newcommand{\Tspc}{\rule{0pt}{2.2ex}}

\begin{document}
\title{\Large \bf Exploiting the Tradeoff between Program Accuracy and Soft-error Resiliency Overhead for Machine Learning Workloads}

\author{Qingchuan Shi, Hamza Omar, Omer Khan \\ 
        University of Connecticut, Storrs, CT, USA}

\maketitle

\thispagestyle{empty}

\setstretch{.98}

\begin{abstract}
To protect multicores from soft-error perturbations, resiliency schemes have been developed with high coverage but high power and performance overheads.
Emerging safety-critical machine learning applications are increasingly being deployed on these platforms.
Moreover, these systems are exposed to harsh environments, such as unmanned aerial vehicles (UAVs)~\cite{Vincent:2013} and self-driving cars~\cite{Junsung:2013}. 
Due to the unique structure and computational behavior of such applications, research has been done on relaxing their accuracy for performance benefits. 
We observe that not all transient errors affect program correctness, some errors only affect program accuracy, i.e., the program completes with certain acceptable deviations from error free outcome.
This paper illustrates the idea of cross-layer soft-error resilience~\cite{shi:cal} using machine learning workloads, where program accuracy is introduced as a tradeoff to deliver resilient yet efficient execution on futuristic large-scale multicores.  
\end{abstract}

\section{Introduction}
The ever increasing miniaturization of semiconductors has led to important advances in mobile, cloud and network computing. However, it has caused electronic devices to become less reliable and microprocessors more susceptible to transient errors. These intermittent faults do not provoke permanent damage, but may result in incorrect execution of programs by altering signal transfers or stored values. These transitory faults are also called soft errors.
As technology continues to scale, industry pundits are projecting that soft-error problem will become increasingly important.
Today's processors implement multicores, featuring diverse set of compute cores and on-board memory sub-systems connected via networks-on-chip and communication protocols. 
Such multicores are widely deployed in numerous environments for their computational capabilities, from traditional applications such as data centers, to emerging areas including unmanned aerial vehicles (UAVs)~\cite{Vincent:2013} and self-driving cars~\cite{Junsung:2013}. 
These cyber-physical systems require high resilience for the safety-criticality~\cite{Augusto:2015,Viguier:2015}, yet high performance for their timing constraints. 
Applications running on such systems include path planning, motion detection, computer vision, and artificial intelligence~\cite{Sermanet:2011,Sato:2014,Haifeng:2012}, where machine learning algorithms are widely used. 
The challenge is to prevent the system running such safety-critical algorithms from failure due to soft-errors, while still meeting the real-time processing constraints.

While extensive research has been done on protecting single processor core from soft-errors, multicore systems introduce new challenges, especially when running parallel applications under complex cache coherence and shared memory protocols~\cite{resilientCoherence:2011}. 
Thread level redundancy (TLR)~\cite{TLR-hpca:2008} is developed to provide high resilience protection for multicore systems. However, such schemes implement coarse-grain checkpoint and roll-back, which introduces high performance overhead.  
A Hardware Redundant Execution (HaRE)~\cite{Shi:2013} scheme was developed earlier, which delivers same protection level as TLR, while avoiding expensive global checkpointing and roll-backs. 
It relies on a local per-core re-execution mechanism to recover from detected errors for the compute core, and implements resilience coherence protocol for the on-chip communication~\cite{resilientCoherence:2011}.

As mentioned earlier, machine learning applications have the potential to be deployed in safety-critical systems, where they face harsh conditions that lead to vulnerability against transient perturbations in the hardware system. 
However, due to the inherent heuristic nature of machine learning algorithms, individual floating point calculations hardly impact program outcome. 
Thus, it is practical to improve processor efficiency by trading off resilience overheads with program accuracy without sacrificing soft-error coverage.
This paper identifies crucial and non-crucial code at program level, based on its impact on the program outcome.
Crucial code affects program correctness, which means the program should be able to complete without crashing, deadlocking, or aborting, etc., due to transient errors, while its outcome is explicable. 
Non-crucial code only affects program accuracy, which refers to how much the result is off compared to no errors scenario.
We have developed a cross-layer resilient architecture~\cite{shi:cal}, through which hardware collaborates with software support to enable  efficient resilience with high error coverage.
Crucial code is executed under HaRE and hence suffers from redundancy's performance overheads.
However, non-crucial code is protected with resilience schemes, which have minimal impact on performance but sufficient to not compromise program correctness.
Only program accuracy is compromised in the worst-case scenario during non-crucial code execution.
This paper evaluates several popular machine learning benchmarks and demonstrate that the cross-layer resilient architecture~\cite{shi:cal} enables significant performance gains while maintaining soft-error coverage.
The proposed scheme incurs $\sim$1.1$\times$ performance overhead over a multicore system without resilience.
In comparison, HaRE incurs a much larger $\sim$1.63$\times$ overhead.

\section{Cross-layer Resilience}
\label{sec:arch}

The key idea of the proposed cross-layer resilience architecture is to protect different code regions with different resiliency schemes.
The novelty comes from the way code regions are identified: based on their impact on program outcome. 
The architecture is applied to trade off performance with accuracy, while maintaining soft-error coverage. 
In order to guarantee high coverage, the architecture uses strong hardware/software protection schemes:
Hardware Redundant Execution scheme~\cite{Shi:2013} (HaRE) for crucial code, and lightweight Software-Hardware Resiliency schemes~\cite{shi:cal} (SHR) for non-crucial code. 
HaRE executes the crucial instruction sequences, which are re-executed and checked to ensure error-free outcome. However, for non-crucial code regions, SHR applies lightweight resiliency schemes that reduce the overhead of redundant executions. 

\subsection{Architectural Support}
\label{sec:hardware}

For crucial code regions, we propose to modify a state-of-the-art hardware resilience mechanism (HaRE) that guarantees high soft-error coverage~\cite{Shi:2013}.
In HaRE, each core re-executes its own atomic instruction sequences, and rollbacks to a safe state when soft-errors are detected. 
For the purpose of deterministic and deadlock-free re-execution, HaRE guarantees atomicity for instruction sequences: modified data is not committed or transferred until control and data flow is checked for soft-errors. 
It has two main phases: \emph{regular execution}, and \emph{redundant execution (re-execution)}. 
At the beginning of regular execution, all necessary states such as register file and program counter are duplicated to ensure a safe state checkpoint.
Re-execution is triggered when the data needs to be consumed, such as when another core sends an invalidation request.
Then the instruction sequence is re-executed and checked to ensure that all the data leaving the core is correct.
Soft-error can be found by comparing a set of control- and data-flow signatures captured during regular and re-execution phases.
In case of mismatch, another re-execution is used to recover from the soft-errors.
A resilient cache coherence protocol~\cite{resilientCoherence:2011} is used to enable holistic coverage for computation and the communication hardware.
Overall, HaRE is able to exploit core level locality for re-execution, and also triggers re-execution on long latency stalls to hide their overhead.
Since HaRE re-executes \emph{all} instructions, its performance can be improved by avoiding it for certain code regions. 
The challenge is to identify code regions that have limited effects to program outcome under soft-error perturbations. 

\subsection{Non-Crucial Instructions}
\label{sec:crucial}

\begin{figure}
\begin{center} 
\includegraphics[height=2in]{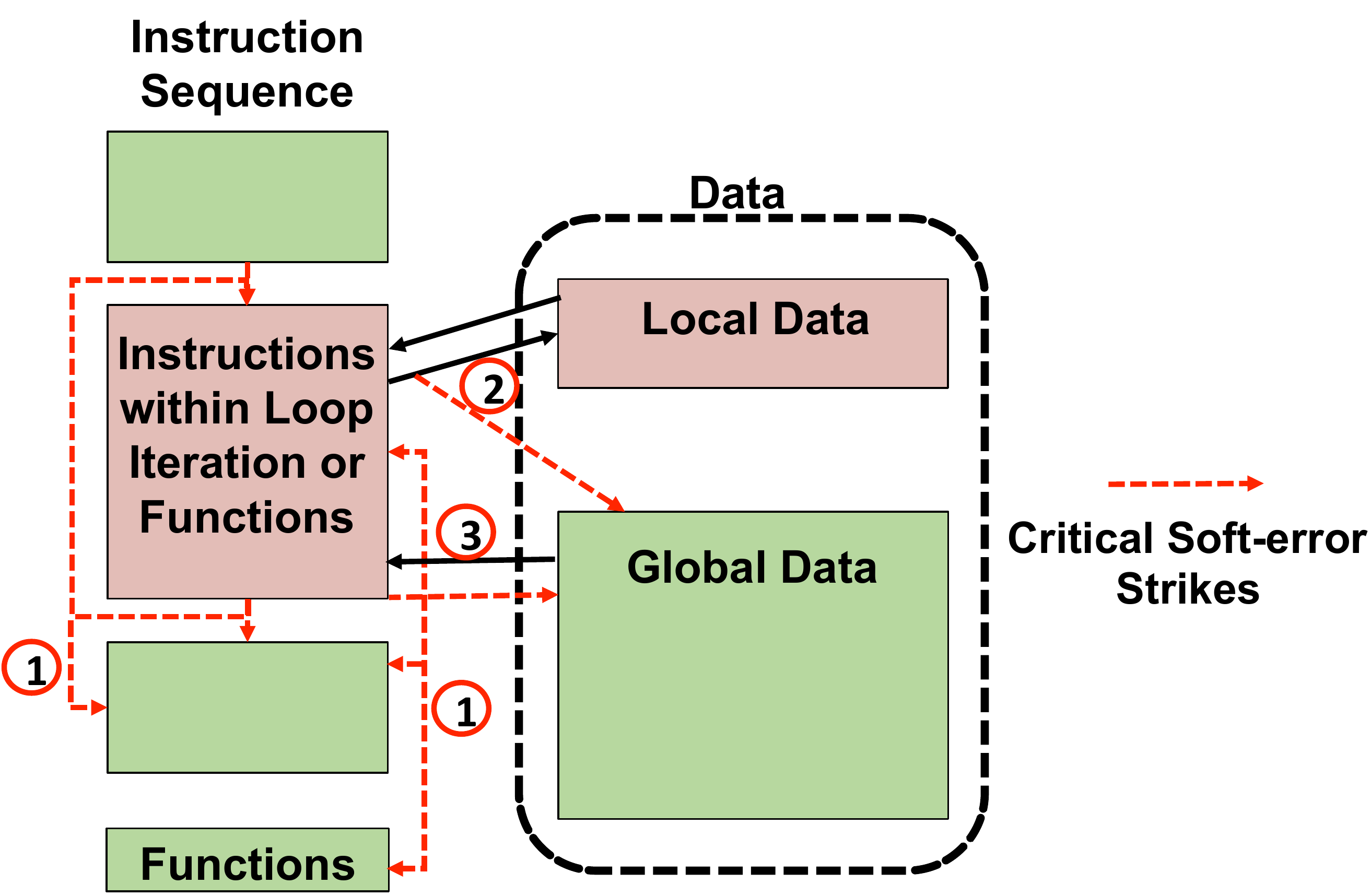}
\caption{Soft-errors' effects to program control and data flows}
\label{fig:softerror} 
\end{center} 
\end{figure}

Soft-errors can affect a program in various ways, such as program crashing, deadlocking, incorrect results and accuracy loss. 
In order to identify the non-crucial code regions, first we analyze the soft-error effects at instruction level.
In cross-layer resilience architecture, all instructions are assumed crucial until proved otherwise.
We consider soft-errors that impact the program control flow or data flow. 

The control flow includes branches, loops, jumps and function calls, as shown in Figure~\ref{fig:softerror}. 
A soft-error can affect the control flow instruction in two ways: wrong target/return address, or wrong branch condition. 
Wrong target/return address can result in accessing arbitrary data and cause unpredictable effects, as shown in Figure~\ref{fig:softerror}:\ding{192}. 
Thus control flow instructions and the calculations of their target/return addresses are always crucial.
Note that when HaRE is enabled around the ``for" statement at program level, calculations of its target/return addresses are automatically protected.
Moreover, a conditional control flow instruction may incorrectly calculate its branch condition under soft-errors.
We rely on the programmer to identify whether the code that resolves the branch target is crucial. 
``for" statements, as well as the updates to loop counter variables are always protected in the applications evaluated in this paper. 
Wrong branch conditions can be tolerated if they do not affect program correctness.

In program data flow, as store instructions directly modify the memory and make the data visible to the program, they are more important than other instructions.
Store instructions can modify the data in arbitrary memory location unexpectedly due to incorrect store addresses, as shown in Figure~\ref{fig:softerror}:\ding{193}. 
Thus store addresses in store instructions, as well as the calculation in case of indirect addressing, are crucial. 
Store instructions can also affect program outcome through data committed to memory. 
In order to illuminate the commit process, the data is divided into two categories: \emph{local} and \emph{global}, as shown in Figure~\ref{fig:softerror}. 
Local data is only used within certain regions. 
It is also temporal because it is not used after exiting the regions, for example, variables defined or initialized within a loop iteration or function. 
Data that is consumed outside the region is considered global. 
Thus local data by itself is non-crucial, and global data can be crucial. 
However local data is accumulated to global data through computations. Values accumulated to global data are considered crucial, as shown in Figure~\ref{fig:softerror}: \ding{194}. Thus computations of local data may also need to be protected to a certain extent for program correctness.

Based on our insights about soft-error effects on program control and data flow, the ideal candidates for non-crucial instructions are compute instructions in-between control flow, which only work on local data and have minimum impact on program outcome. 
One example would be random number generation instructions in Monte Carlo method. 
However, in real applications, instructions that meet all these conditions are rare.
Thus in the next section, we introduce lightweight resilience schemes to include more instructions in non-crucial code regions. 

\subsection{Composing Non-Crucial Code Regions with SHR}

\begin{figure}
\begin{center}
\includegraphics[width=1.0\linewidth]{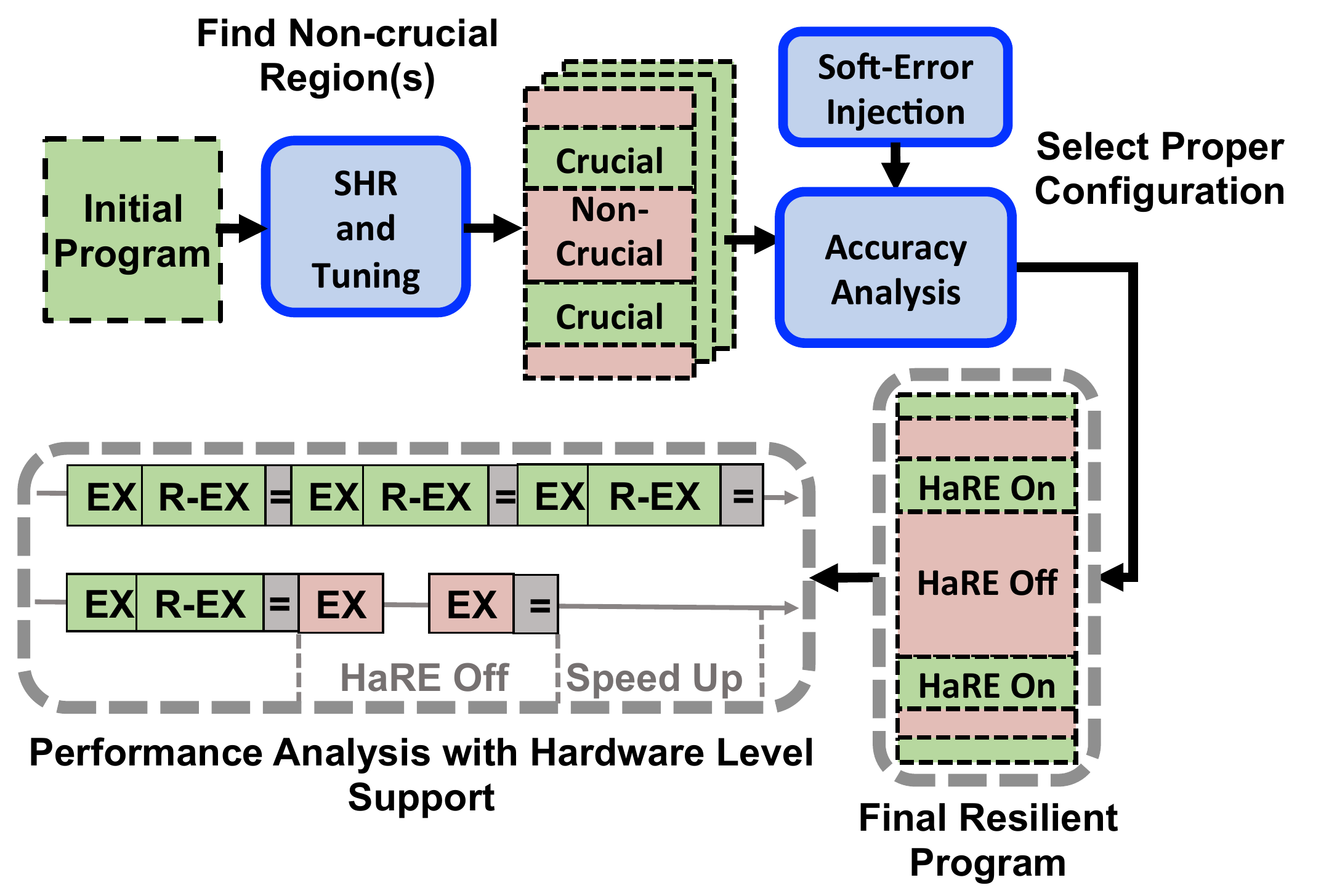}
\caption{Cross-layer Resilience Framework Workflow.}
\label{fig:flow}
\end{center}
\end{figure}

In addition to the fact that naturally non-crucial instructions are rare, turning on/off HaRE introduces performance overheads.
In order to make the architecture beneficial for the programs, more instructions need to be made resilient for certain soft-errors, and it is favorable to have them in continuous sequences. 
In other words, the soft-error effect to the program outcome needs to be restricted for certain code regions.
For this purpose, Software-Hardware Resiliency schemes (SHR) are introduced.

First, at the hardware level, SHR applies protection to store instructions. 
Based on the insight that store addresses are always critical to the program, SHR performs hardware-level redundant address calculations with additional overheads. Store operations proceed after their address calculation is verified. Otherwise they are re-executed.
Second, at the software level, the value committed to global data is checked when necessary, to ensure program correctness and maintain accuracy. 
In machine learning applications, majority of critical variables are bound-checkable. 
Thus it is practical to use a programmer defined, software level bound checker to provide certain resilience protection.

With SHR, store instructions and local data computations with predictable outcome are able to be included in non-crucial code regions.
These make it more feasible to form individual non-crucial instructions into code regions.
When taking SHR into consideration, code regions can be classified as non-crucial if they do not contain control flow instructions, and  meet one of the following: (1). They do not access global data; (2). The value committed to global data does not affect the program outcome. (3). The value committed to global data can be bound checked.
Based on the relaxed criteria, a reasonable amount of code in machine learning applications can be considered non-crucial.
Schemes such as software level bound checking and loop-unrolling are used to compose non-crucial instructions into code regions. 
This provides better performance while ensuring coverage.

\subsection{Selecting Non-crucial Code Regions with Acceptable Accuracy}
\label{sec:accuracy}

Figure~\ref{fig:flow} shows the design flow, where the programmer first identifies all candidate non-crucial code regions.  
Accuracy analysis is performed to verify the accuracy loss of the non-crucial code regions, and help the programmer select the proper combination of regions under certain constraints, such as soft-error rate and program accuracy loss threshold.
This combination is a subset of all the non-crucial regions, which is referred as configuration in this paper.
With the proper configuration, the transformed program can be deployed on the proposed cross-layer architecture.
In the current setup, programmer's effort is mainly spent in non-crucial code identification. 
The other steps are assisted and automated to a certain extend. 

\begin{figure}
\begin{center}
\includegraphics[width=0.8\linewidth]{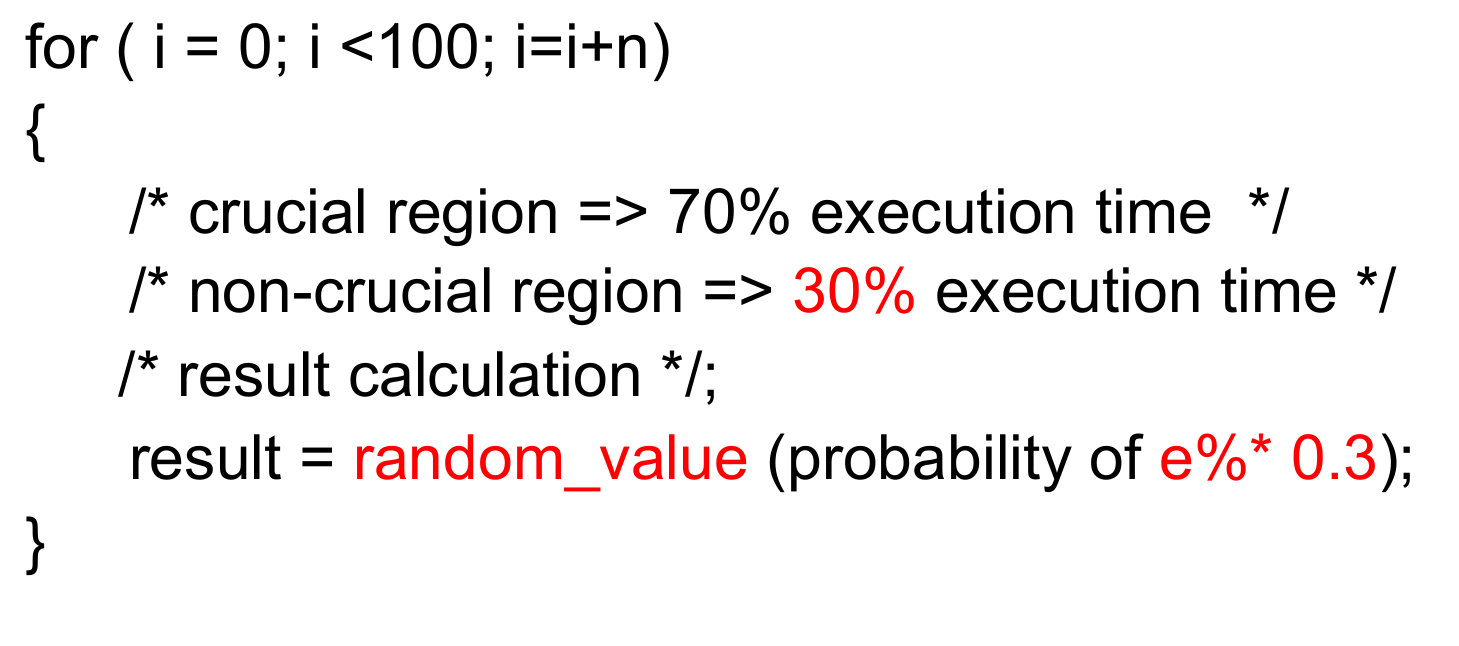}
\caption{Example of fault injection flow for a loop.}
\label{fig:fault_injection}
\end{center}
\end{figure}

In order to obtain the program accuracy, the accuracy loss of each non-crucial code region is obtained by performing program level fault injection experiments.
It imitates random soft errors in non-crucial code regions.
It is done in such a way that a program variable prone to soft-errors is exposed to random values in the range determined by its data type.
The probability of the injection is determined based on the error rate ($e\%$), and the execution time of that non-crucial code.
As shown in Figure~\ref{fig:fault_injection}, when assuming an error rate of $e\%$ and 30\% of the code (based on execution time) is non-crucial in each iteration, the probability of perturbing the result of single iteration would be $e\% \times 0.3$.  
The notion of the fault injection is that after applying the protection schemes, the resilience framework ensures store address and control flow instructions are always protected (using SHR or HaRE). 
The remaining vulnerable code is other data flow instructions, meanwhile the out of bound value is never committed.

The accuracy loss of each region is obtained by only applying fault injection to it, where the rate of ``correct classification divided by the number of tests" is defined as the accuracy.  
For example, when applying 100 hand written digits through CNN-MNIST workload, if 95 digits are classified correctly, its accuracy is defined as 95\%. 
Multiple simulations (1000 times in this paper) are performed for each configuration of non-crucial regions to obtain the average program accuracy.
The aggregate accuracy loss due to errors in non-crucial regions depend on a programmer defined accuracy threshold, as well as the error rate due to system and environmental conditions.

In order to select the proper configuration under given constraints, all non-crucial code regions are included in the configuration at the beginning. 
If the program accuracy loss exceeds the threshold, the region with maximum accuracy loss in the previous configuration is considered as crucial.
In case multiple regions have similar accuracy loss, the workflow chooses the one with minimum execution time.
In the worst case, all non-crucial code regions are considered crucial.

As shown in Figure~\ref{fig:flow}, after selecting the proper configuration, switching between crucial and non-crucial is placed in the program. 
These are passed as HaRE on/off pragmas to the hardware. 
The hardware-software interface is implemented using a special function instrumented in the programs.
HaRE is turned on before the bound checker to protect it.
The final resilient program is executed on the proposed cross-layer architecture, with hardware level support implemented in the simulator.
Detailed code demonstrations are shown next.

\section{Application Illustration}
\label{sec:application}

Machine learning applications are ubiquitously used in many domains, where the systems could be exposed to soft-errors.
Due to their unique structure and computational behaviors, research has been done on relaxing their accuracy for performance benefits. 
Likewise they have potential in the proposed architecture for resilience--accuracy tradeoff. 
We evaluate six machine learning benchmarks.

\subsection{Convolutional Neural Networks (CNN)}

\begin{algorithm}
\caption{CNN Convolutional Layer Pseudo Code}
\begin{algorithmic}[1] 
\State \textbf{ConvolutionLayer\emph{(input, conv\_out, tid, threads)}} \{
\For{each neuron in the thread}
\State $\backslash*$ The following 3 level loop is Unrolled $*\backslash$
\For {(num. of kernels $k$, kernel height $h$, kernel width $w$)} 
\State $\backslash*$ Assign temp\_k/h/w $*\backslash$
\State \emph{\textcolor{red}{HaRE Off}} 
\State $conv\_out$ += $do\_conv(input,temp\_k/h/w)$ 
\State $\backslash*$ Update temp variables $*\backslash$
\State $conv\_out$ += $do\_conv(input,temp\_k/h/w)$ 
\State $\backslash*$ Update temp variables $*\backslash$
\State :
\State \emph{\textcolor{red}{HaRE On}}  
\State $\backslash*$ Update k, h, w $*\backslash$
\EndFor
\State $Bound\_Checker(conv\_out)$
\EndFor
\State \}
\end{algorithmic}  
\label{alg:cnn-conv}
\end{algorithm} 

CNN is a highly prevalent type of neural network.
This paper evaluates four of the most commonly used convolutional neural networks: 
AlexNet (ALEXNET)~\cite{krizhevsky2012imagenet}, hand written digits recognition (MNIST)~\cite{Michael:mnist}, recognition of German traffic signs (GTSRB)~\cite{Sermanet:2011}, and VGG~\cite{vgg}.
All of them consist of 4 types of layers: input, convolutional, fully connected, and output layers.
From a programmer standpoint, the computation within each layer can be identified as non-crucial.
In this section convolutional and fully connected layers are illustrated, as they contribute most of the execution time.
The parallelization strategy for CNN is to divide the neurons in each layer among the available threads. 
Since the subsequent layers consume the outputs of prior ones, barriers are used to synchronize the threads after each layer, which are protected through HaRE.

In CNN, the convolutional layer takes the input feature map, then convolves it with the kernels to give an output feature map. 
This results in an output feature matrix cell value. 
These computations (shown in Algorithm~\ref{alg:cnn-conv} lines 7-11) can be considered as non-crucial, since the effect of individual cell value to the program outcome is limited. 
Furthermore, each kernel produces an output feature matrix of its own.
Cell values are compared with each other to find the maximum one, which is later used to construct a single output feature map. 
When exposed to soft-errors, the output map can get affected only if the maximum cell values are perturbed into larger ones, since smaller values are masked out. 
Note that out of bound values are dropped. 
In that case, the second largest value would be used for the corresponding cell in the output feature map. 
The loops (shown in Algorithm~\ref{alg:cnn-conv} line 4) are unrolled, and the loop counters (k, h, w) are updated with HaRE protection. Only temporary variables (temp\_k/h/j in Algorithm~\ref{alg:cnn-conv}) are updated in non-crucial region.

\begin{algorithm}
\caption{CNN Fully Connected Layer Pseudo Code}
\begin{algorithmic}[1] 
\State \textbf{FullyConnectedLayer\emph{(input, fully\_out, tid, threads)}} \{
\For {each layer} 
\For{each neuron}   
\State $\backslash*$ The following loop is Unrolled $*\backslash$
\For {each input $i$}   
\State \emph{\textcolor{red}{HaRE Off}} 
\State $O$ += $(input(temp\_i) * weights(temp\_i))$ 
\State $temp\_i$ = 1
\State $O$ += $(input(temp\_i) * weights(temp\_i))$ 
\State $temp\_i$ = 2
\State :
\State \emph{\textcolor{red}{HaRE On}}
\State $\backslash*$ Update i $*\backslash$
\State $Bound\_Checker(O)$
\EndFor
\State $fully\_out$ $=$ $Sigmoid(O)$
\EndFor
\State \emph{\textbf{Barrier}}
\EndFor
\State \}
\end{algorithmic}  
\label{alg:cnn-fully}
\end{algorithm} 

The fully connected layer (as shown in Algorithm~\ref{alg:cnn-fully}) in CNN is a feed-forward network, in which all the neurons in one layer are connected to the neurons in the next layer. 
The neuron count reduces towards the end of a fully connected layer. 
The first layer of this feed-forward network provides the output data set of the previous layer to the neurons as input. 
The later layers perform accumulations and multiplications of the inputs with their respective weights to compute the sigmoid. 
The result is further propagated to the next layer. 
The computations done in the fully connected layer can be considered non-crucial, since the remaining unperturbed accumulations could overcome it.
To ensure correctness of the program, bound checkers (shown in Algorithm~\ref{alg:cnn-conv} line 14 and Algorithm~\ref{alg:cnn-fully} line 14) are introduced in the code so that the effects of the perturbations can be reduced. Statically determined bounds are used. 
For fully connected layers, the accumulated results are used to compute the sigmoid. 
Based on the definition, this complex sigmoid computation always results in a value within the range of 0 to 1. 
For the sigmoid to output a 0 or a 1, the respective values (\emph{O}) are -90 and 10. 
Thus, to limit the result with in this range of 0 to 1 (excluding 0 and 1), the accumulations are limited to -90 and 10. 

\subsection{Multi-Layer Perceptron (MLP)}
Multi layer perceptron (MLP) is a feed-forward neural network in which all the neurons in one layer are connected to every neuron in the next layer. 
The structure of MLP resembles the fully connected layer of CNN (Algorithm~\ref{alg:cnn-fully}). 
There are three types of layers in MLP namely input layer, intermediate layer(s) and an output layer. 
Barriers are implemented between the layers. 
The input layer of MLP feeds input data set to neurons. 
If it is perturbed, neurons will work on wrong initial data, and propagate the errors.  
Thus the input layer is always considered crucial. 
The calculation in intermediate layers can be considered non-crucial.
Similar to the fully connected layer of CNN, bound checkers are applied for sigmoid calculation.

\begin{algorithm}
 \caption{Resilient KNN Pseudo Code}
 \label{alg:knn}
 \begin{algorithmic}[1]
 \State \emph{\textcolor{red}{HaRE On}} 
 \For {(each sample, $s$)}              
 \For{(naive sample, $n$)} 
 \State \emph{\textcolor{red}{HaRE Off}} 
 \State \emph{$D$ = Calculate\_Distance$($s$,$ n$)$} \Comment{Unrolled}
 \State \emph{\textcolor{red}{HaRE On}}
 \State \emph{Sort $($D$)$}
 \EndFor
 \EndFor                
 \end{algorithmic}
\end{algorithm}

\subsection{K-Nearest Neighbors (KNN)}
In KNN, objects are classified using a number of known examples called training data.
The distances between the new object and each known object is calculated and are sorted to determine k-nearest neighbors.
Then, the class of the new object is decided by majority vote over k-nearest neighbors.
The outer loop (Algorithm~\ref{alg:knn} line 2) of KNN visits all the test samples while the inner loop  (Algorithm~\ref{alg:knn} line 3) traverses over all the training data (naive samples). KNN uses inner loop parallelization where each thread calculates the euclidean distances between the known and test samples, and sorts them.

The Calculate\_Distance function (Algorithm~\ref{alg:knn} line 5) in KNN involves accumulations of large amount of examples. Each example has minimum impact on the program, hence its calculation can be considered as non-crucial. 
If a distance value gets perturbed, its affect on the overall outcome would be insignificant. Moreover, as the distance calculation of one neighbor is independent of the other neighbor, the perturbations do not propagate.

\section{Evaluation}
\begin{figure}
\begin{center}
\includegraphics[width=1.0\linewidth]{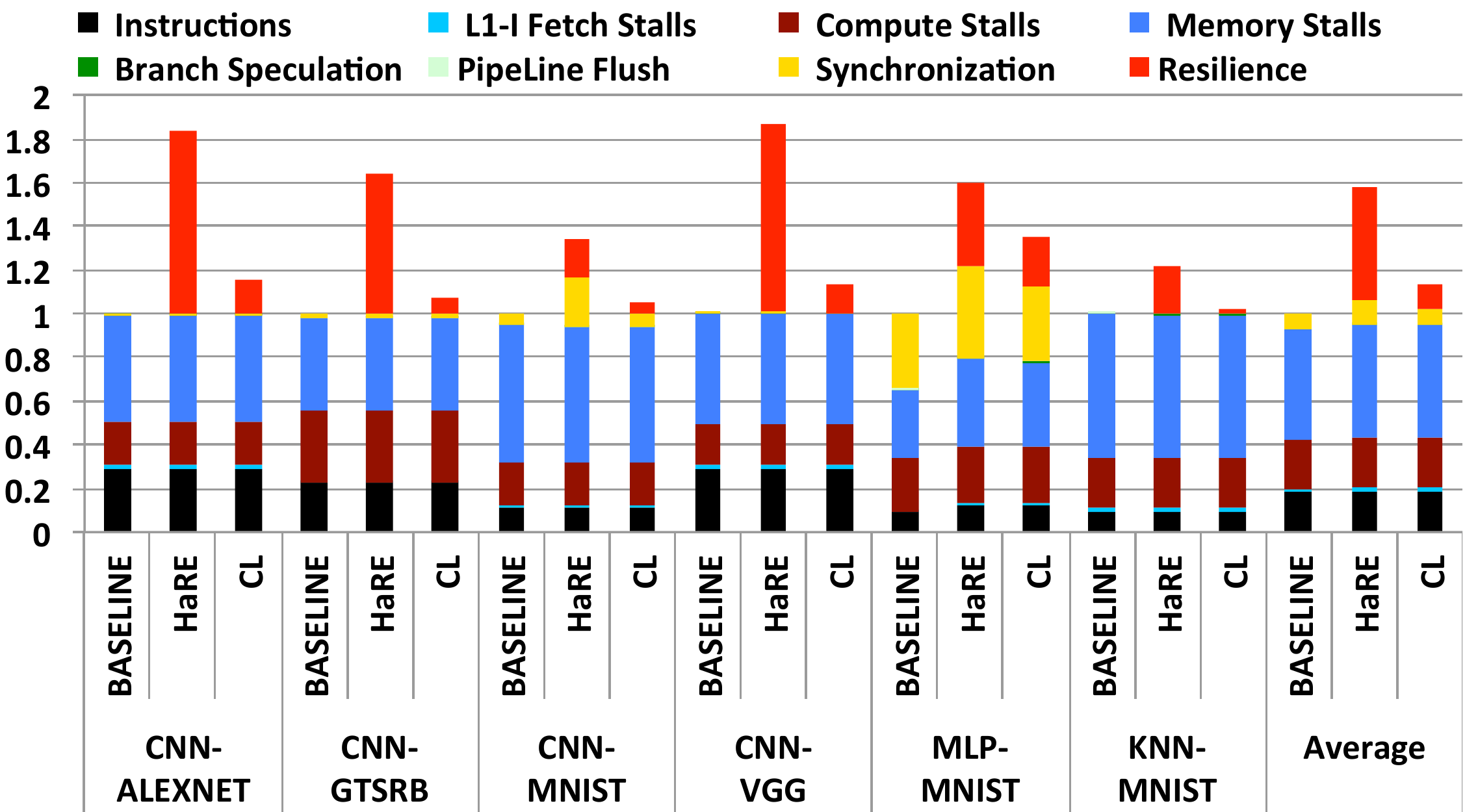}
\caption{ Completion time of selected configurations using proposed cross-layer architecture at 0.1\% error rate and 10\% accuracy threshold.}
\label{fig:result}
\end{center}
\end{figure}

We modify the Graphite multicore simulator~\cite{Miller:2010} to evaluate the proposed cross-layer resilient architecture.
The default architecture parameters are summarized in Table~\ref{tab:architectural_parameters}.
Results in Figure~\ref{fig:result} are normalized to BASELINE multicore system with no resilience protection schemes.
HaRE redundantly executes all instructions. CL is the proposed cross-layer architecture, which selectively applies HaRE and SHR.

\begin{table}
\centering
\footnotesize
\begin{tabular}{|l|p{4.5cm}|}
\hline
   Architectural Parameter & Value           \\
   \hline \hline
   Core                     & 64 In-Order, Single-Issue cores           \\
   \hline \hline
   \multicolumn{2}{|c|}{Memory Subsystem}                      \\
   \hline
   L1-I/D Private Caches                & 32~KB, 4-way Set Assoc., \\
   (per Core)                           & 1 cycle latency\\
   L2 Shared Cache                      & 256 KB, 8-way Set Assoc., \\					
   (per Core)                           & 8 cycle latency, Inclusive \\					
   Coherence Protocol				    & Directory, Invalidation-based, MESI \\
   DRAM Memory Interface		        & 8~controllers, 5~GBps/controller, 100~ns latency                \\
   \hline \hline
   \multicolumn{2}{|c|}{Electrical 2-D Mesh with XY Routing}                       \\
   \hline
   Hop Latency                         & 2 cycles (1-router, 1-link) $+$ link contention \\
   Flit Width                                & 64 bits                  \\
   \hline
\end{tabular}
\caption{Architectural Parameters.}
\label{tab:architectural_parameters}
\end{table}

\subsection{Performance}  

First, we evaluate the performance of the proposed cross-layer architecture. 
Under normal conditions, the probability of soft-error strikes is very low (less than 1 per day for the current technology node~\cite{Premkishore:2002}).
In order to reveal the soft-error effects to program accuracy, an aggressive error rate of 0.1\% is used.
We define the accuracy bound as 90\% to select the acceptable program configuration. 
The completion time of selected configuration of each benchmark is plotted in Figure~\ref{fig:result}.

HaRE performs reasonably well for most benchmarks, because it performs local redundant execution and exploit core-level locality.
Moreover, it hides re-execution latency behind cache miss stalls. 
Memory stalls of HaRE are increased over BASELINE, because memory operations trigger re-executions (such as invalidating a cache line). 
Messages need to wait until the re-execution completes. 
The synchronization of HaRE also increases over BASELINE due to the re-execution time of instructions within locks or barriers. 

\begin{table}
\centering
\footnotesize
\begin{tabular}{|l|l|l|l|l|}
\hline
\Tspc \textbf{}  & Selected    & Non & Accuracy  \\
\Tspc            & Non-crucial &  -Crucial  & Loss (\%) \\
\Tspc            & Regions     &  Time (\%)           & \\
   \hline 
   \hline
 \Tspc {CNN-VGG}  &  {C: 1-16 + F: 1-3} &  {92.5} &   {7.3}\\                                 
   \hline
 \Tspc {CNN-}  &  {C: 1-5 + F: 1-3} &  {91} &   {6.9}\\                                 
 \Tspc {ALEXNET}  &   &   &   \\                                 
   \hline
\Tspc {CNN-GTSRB}    &  {C: 1,2 + F: 1,2} &  {88} &  {6.2}  \\
 \hline    
\Tspc {CNN-MNIST}    &  {C + F} &  {87} &   {3.9} \\
   \hline   
   \Tspc {MLP-MNIST}    &  {I: 1,2} &  {75} &  {3}  \\                                      
   \hline
   \Tspc {KNN-MNIST}    &  {Distance} &  {94} &  {1.7}  \\
\hline
\end{tabular}
\caption{Selected configurations of machine learning benchmarks at 0.1\% error rate and 10\% accuracy threshold. (C - Convolutional Layer, F - Fully-Connected Layer, I - Intermiediate Layer)}
\label{tab:mlselectconfigs}
\end{table}

When applying CL, all machine learning benchmarks show remarkable performance improvement over HaRE. 
CL reduces completion time significantly compared to HaRE (from 1.83$\times$ to 1.15$\times$ for CNN-ALEXNET). 
This is because HaRE is not able to hide resilience overhead, meanwhile a major amount of computations are identified as non-crucial in CL.
Overall, the proposed CL architecture shows significant performance improvement over HaRE. It reduces the performance overhead of resiliency from $\sim$1.63$\times$ to $\sim$1.1$\times$ on average.

\subsection{Accuracy} 
Table~\ref{tab:mlselectconfigs} shows the percentage of time spent in non-crucial code and the percentage of accuracy loss for the selected configurations at an error rate of 0.1\%.
Overall, CL is able to select configurations with reasonable amount of non-crucial code, resulting in low accuracy loss.
The accuracy loss in different benchmarks highly depends on the code structure and the functionality.  
For code regions with similar functionality in a benchmark, the more execution time the code region has, the more accuracy loss it may contribute. This behavior is observed in the convolutional layers of CNN.
However, this pretext would not hold if the code regions have different functionalities, such as the output layer in all machine learning benchmarks. 
It does not have much execution time, but contributes a relatively large amount of accuracy loss.
This is because errors in the output layer could directly affect the final program outcome.
In fact, all machine learning benchmarks have more than 10\% accuracy loss at 0.1\% error rate, if their output layers are defined as non-crucial.

\subsection{Configuration Selection} 

\begin{figure}
\begin{center}
\includegraphics[width=0.9\linewidth]{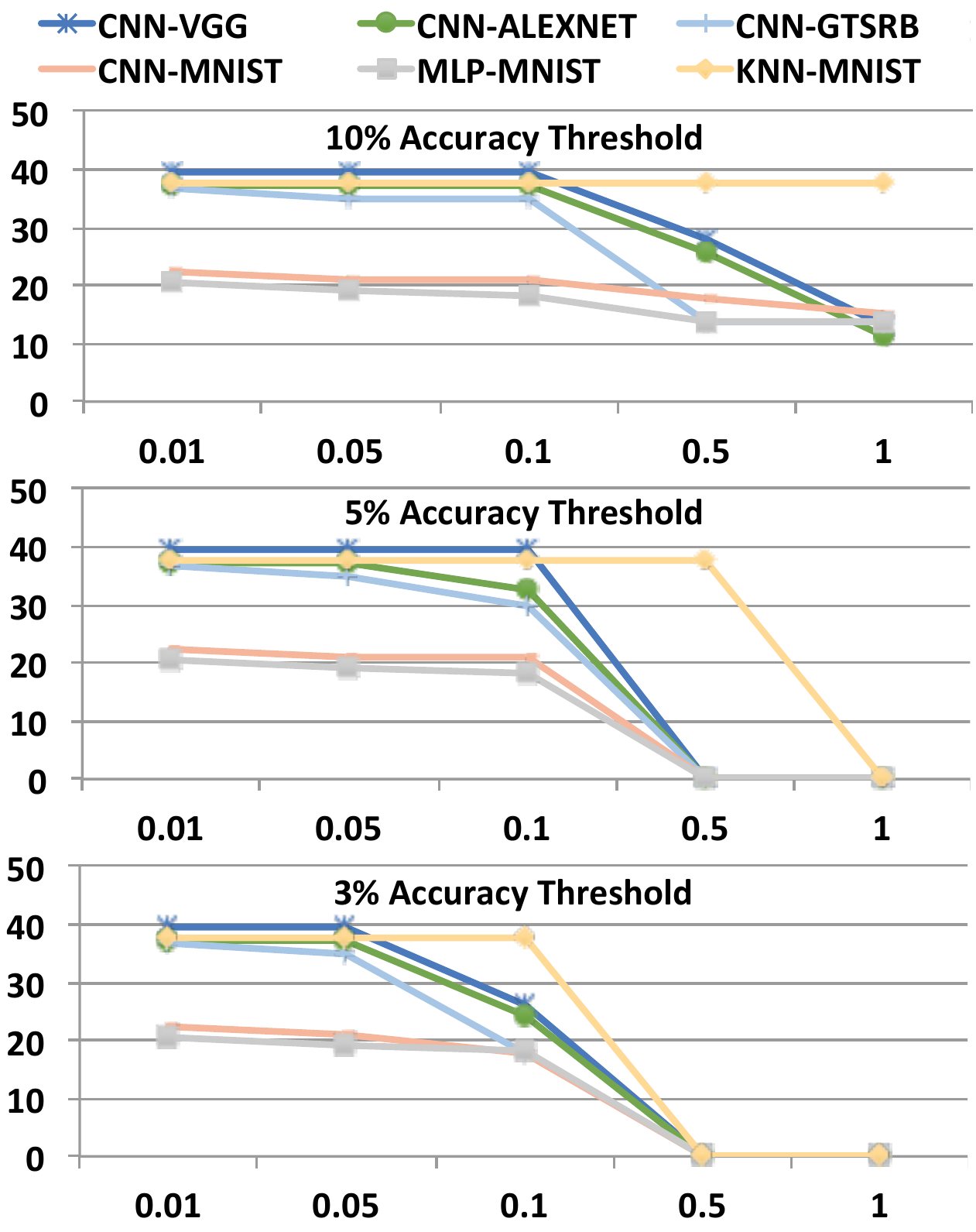}
\caption{CL \% performance improvement (Y axis) over HaRE at 10\%, 5\% and 3\% accuracy thresholds with error rates from 0.01\% to 1\% (X axis).}
\label{fig:acc_perc}
\end{center}
\end{figure}

Soft-error rates change due to different conditions in real world environmental conditions. 
In addition, the acceptable accuracy loss is determined from case to case. 
In Section~\ref{sec:accuracy}, we introduced how to select the proper configuration. 
Figure~\ref{fig:acc_perc} shows the performance improvements over HaRE of selected configurations, at different accuracy thresholds with different error rates. 

As shown in Figure~\ref{fig:acc_perc}, three different accuracy thresholds (10\%, 5\%, and 3\%) are applied, from top to bottom.
In general, a higher accuracy threshold and lower error rate results in configurations with more non-crucial code regions. 
For example, when using an accuracy threshold of 10\%, CNN-ALEXNET is able get the best performance at 0.1\% or lower error rates.
According to Table~\ref{tab:mlselectconfigs}, all its convolutional and fully connected layers are considered as non-crucial at this point, which contribute 91\% of the program's execution time. 
Thus, the proposed cross-layer architecture has great performance improvement over HaRE (37\%).  
As the error rate increases, the same configuration cannot hold a 10\% accuracy threshold. 
It gets to 31\% at an error rate of 0.5\%, which was 7\% earlier (error rate of 0.1\%).
We are able to get the accuracy loss back by making the 5th convolutional layer and the 3rd fully connected layer crucial.
However, this results in less non-crucial code (76\% of execution time), thus has less performance improvement (26\%).
In extreme cases, the performance improvement can reduce to 0\%. This means all code regions have to be considered as crucial, which is effectively program execution under HaRE for all instructions. 

\section{Conclusion}
\label{sec:conclusion}

This paper introduces a novel cross-layer resilience architecture, and illustrates it with machine learning applications.
The key idea is to explore resilience overhead tradeoff with program accuracy, while not compromising soft-error coverage and safe execution of the program.
It guarantees program correctness, while only incurs $\sim$1.1$\times$ performance overhead over a system without resilience.
This is an average 32\% performance improvement over state-of-the-art hardware resilience scheme that protects the whole program.

\setstretch{1}

\bstctlcite{bstctl:etal, bstctl:nodash, bstctl:simpurl}
\bibliographystyle{ieeetr}
\bibliography{refs}

\end{document}